# Quasi-TEM modes in rectangular waveguides:
## a study based on the properties of PMC and hard surfaces

**R. Pierre, G. Tayeb\*, B. Gralak, S. Enoch**

Institut Fresnel, UMR CNRS 6133, Université Paul Cézanne Aix-Marseille III
Faculté des Sciences et Techniques, 13397 Marseille Cedex 20, France
*Corresponding author. Email: gerard.tayeb@fresnel.fr

**Abstract**
Hard surfaces or magnetic surfaces can be used to propagate quasi-TEM modes inside closed waveguides. The interesting feature of these modes is an almost uniform field distribution inside the waveguide. But the mechanisms governing how these surfaces act, how they can be characterized, and further how the modes propagate are not detailed in the literature. In this paper, we try to answer these questions. We give some basic rules that govern the propagation of the quasi-TEM modes, and show that many of their characteristics (i.e. their dispersion curves) can be deduced from the simple analysis of the reflection properties of the involved surfaces.

Keywords: TEM waveguide, quasi-TEM mode, hard surface, perfect magnetic conductor (PMC).

## 1. Introduction

Quasi-TEM modes with an almost uniform electric field repartition are interesting in many domains. For instance, they can be used for resonant cavities applications [1], horn antennas as well as waveguides [2-5]. Similarly to some recent papers [6,7], the main issue of our study is the design of resonant cavity antennas with the goal of improving their aperture efficiency. Standard Perfect Electric Conducting (PEC) enclosures do not allow the propagation of quasi-TEM modes since they force the tangential electric field to vanish on it and the required boundary condition (non vanishing tangential electric field) not to be fulfilled. Consequently, these modes require the waveguide walls to be replaced by perfect magnetic conducting surfaces (PMC) or hard surfaces [8-10]. While implementing these solutions, we were led to many interrogations and their answers were not clearly given in the literature. For instance, a quick analysis of the problem seems to show that the problem is strongly linked with the reflection of a plane wave for grazing incidence. But many of the studies use normal incidence in order to characterize the phase of the reflection coefficient of these surfaces... Another problem is to choose between the numerous surfaces available: PMC or hard? And between the numerous possibilities in order to build these surfaces: PEC substrate loaded with a dielectric slab [1,11], corrugated surfaces [5,11], FSS [4], dielectric slab covered with strips or dipoles [4,7,12].

The study of these structures can be very cumbersome, using 3D numerical codes (modeling cavities) or 2D ones (modeling waveguides). Since our group has been from a long time specialized in the study of diffraction gratings (periodic 1D surfaces), it was interesting to investigate the properties of PMC and hard surfaces using these fast 1D codes [13-15]. We show hereafter that it allows us to characterize their fundamental behavior, and predict some properties of the quasi-TEM modes that may propagate in waveguides build from these surfaces. By these means, we can access interesting information about the modes, and their dispersion curves. We can also answer questions about how do these surfaces behave: it is shown that the convenient boundary conditions are obviously not fulfilled on the surface itself, but in a close vicinity to the surface, thanks to the presence of evanescent modes.

We illustrate our results with the study of a square waveguide with four hard walls that allow a dual-polarized quasi-TEM mode.

## 2. Intuitive TEM mode propagation in grazing incidence

The most intuitive conditions that could allow the propagation of TEM modes inside a rectangular waveguide are depicted in Fig. 1. We can think to a "xy-truncated" plane wave propagating along the z-axis (the waveguide invariance direction), in such a way that the waveguide walls allow the appropriate boundary conditions: the electric field should be normal to surfaces 1 and 3, and tangential to surfaces 2 and 4.

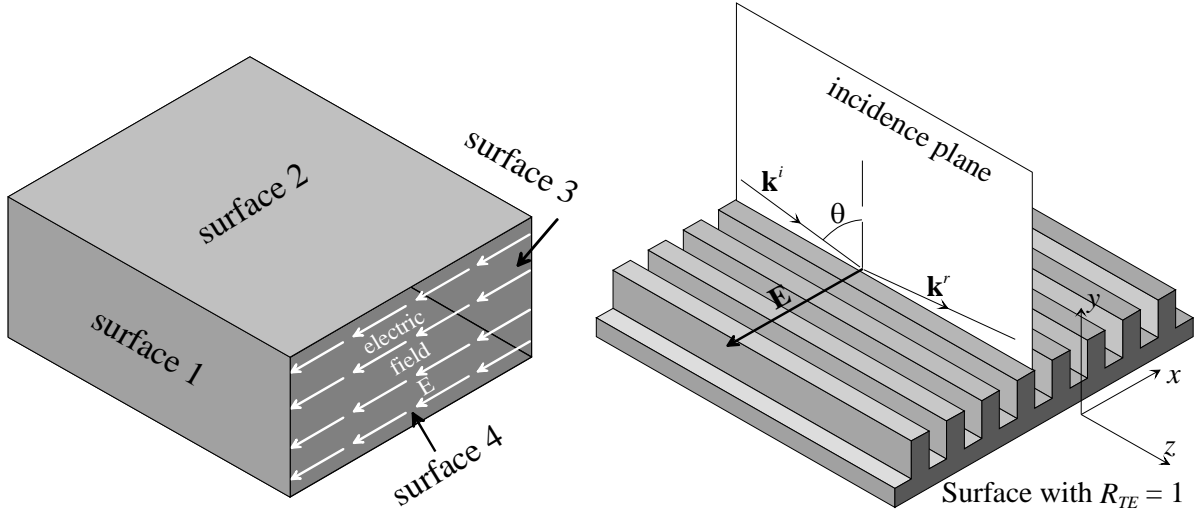

*Fig. 1: A TEM mode propagating along z inside a rectangular waveguide (left) and a TE-polarized plane wave reflected at grazing incidence ($\theta \approx 90°$) on a surface used as surfaces 2 and 4 (right).*

Table 1 provides some properties of the PEC, PMC, soft and hard surfaces in terms of reflection coefficient of a plane wave in TE (resp. TM) polarization, i.e. when the electric (resp. magnetic) field is perpendicular to the plane of incidence (see [9] for more details). We suppose that the considered surfaces are lossless. In the case of a periodic surface, we suppose also that there is only one reflected order. It implies that the modulus of the reflection coefficient is equal to 1. In order to get the TEM mode, the electric field should not vanish on surfaces 2 and 4 that should be either PMC or hard for grazing incidence ($\theta \approx 90°$). As regards surfaces 1 and 3, PEC are obviously possible, but also hard surfaces, since they allow a non-vanishing tangential magnetic field. Note that if all four surfaces are hard, the propagation of a TEM mode with orthogonal polarization (E field parallel to the *y*-axis) is also possible.

| Surface | Reflection coefficients | Boundary field components |
|---|---|---|
| PEC | $R_{TE} = -1$ | $E_t = 0$ |
| | $R_{TM} = +1$ | $H_t \neq 0$ |
| PMC | $R_{TE} = +1$ | $E_t \neq 0$ |
| | $R_{TM} = -1$ | $H_t = 0$ |
| Soft surface | $R_{TE} = -1$ | $E_t = 0$ |
| | $R_{TM} = -1$ | $H_t = 0$ |
| Hard surface | $R_{TE} = +1$ | $E_t \neq 0$ |
| | $R_{TM} = +1$ | $H_t \neq 0$ |

*Table 1: Properties of PEC, PMC, soft and hard surfaces concerning the reflection coefficient of a plane wave and the associated conditions on the fields in the vicinity of the surface.*

### 3. Study of surfaces using reflection coefficients

Following the previous section, it is interesting to study some realizations which have been proposed for the making of PMC and hard surfaces. We restrict ourselves to 1D periodic surfaces (periodicity along the x-axis, invariance along the z-axis, according to Fig. 1), since we have developed fast numerical codes able to handle such structures. We investigate the reflection coefficient of a plane wave impinging an infinite surface in the conditions of Fig. 1, i.e. the plane of incidence is the $yz$-plane, with an incidence θ, and for the two fundamental polarization cases (TE and TM). The incident plane wave has unitary amplitude. We plot the phase of the reflection coefficients (the argument of the complex numbers $R_{TE}$ and $R_{TM}$ ) versus the wavelength λ and the angle of incidence θ. In the ideal situation where $R_{TE} = 1$ and for grazing incidences, the incident and the reflected plane waves should add themselves, and give a tangential electric field on the surface with amplitude equal to 2. Since the surfaces have metallic parts, we are conducted to an apparent contradictory situation, since the metallic surface imposes the vanishing of the tangential electric field. But one must remember that the field is not the only superposition of the incident and reflected plane waves. There are also evanescent waves in the vicinity of the surface, and they play a vital role in the realization of the boundary conditions. That will appear clearly hereafter.

Let us start with the simpler surface used to get PMC boundary conditions, i.e. a grounded slab, which is a metallic ground plane covered by a dielectric layer. Denoting by $h$ and ε the thickness and the relative permittivity of the layer, one shows that $R_{TE} = +1$ is obtained when the following condition is realized:

$$h = \frac{\lambda}{4\sqrt{\varepsilon - \sin^2\theta}} \, . \tag{1}$$

Figure 2 shows the phase of $R_{TE}$ and $R_{TM}$ for a grounded slab with $\varepsilon = 2.2$ and $h = 4\,\text{mm}$. Along the dashed curve representing Eq. (1), we have simultaneously $R_{TE} = +1$ and $R_{TM} = -1$, as expected for a PMC surface. For our purpose, we mostly focus on the condition $R_{TE} = +1$, i.e. the region close to the dashed curve on the Fig. 2 (left). We can see that near the interesting point concerning the propagation of the TEM mode described in the previous section (obtained for $\theta \approx 90°$ and $\lambda \approx 17.5\,\text{mm}$), the bandwidth is quite small.

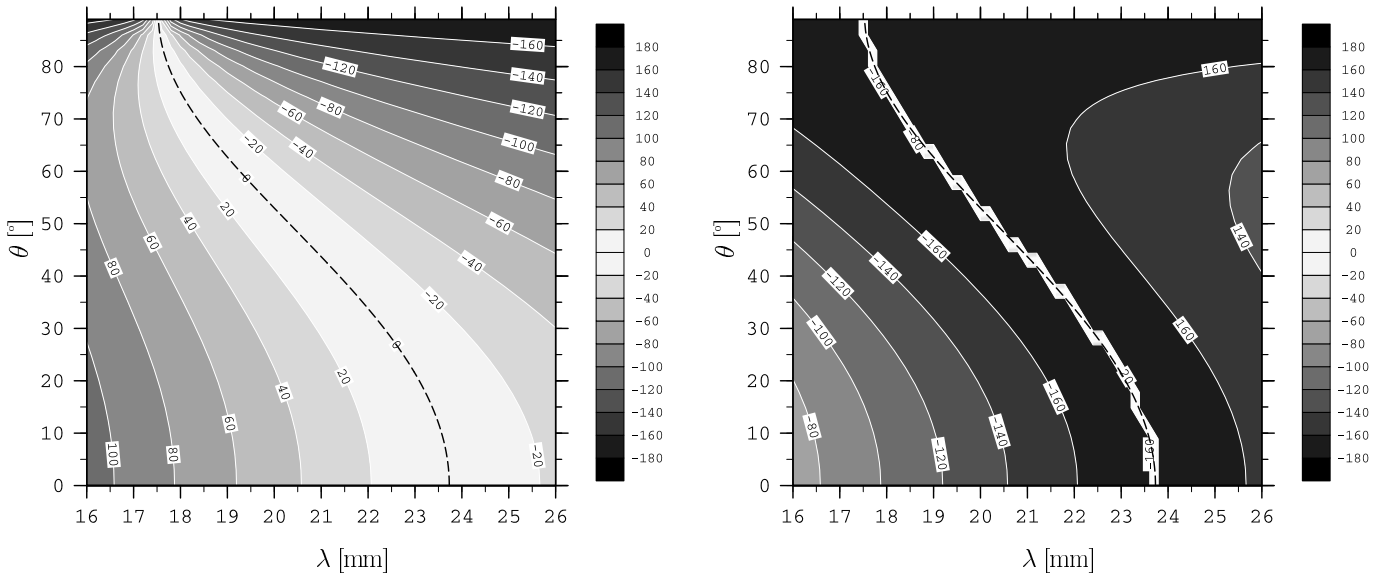

*Fig. 2: Phases (in degrees) of $R_{TE}$ (left) and $R_{TM}$ (right) for a grounded slab with $h = 4\,\text{mm}$ and $\varepsilon = 2.2$. The dashed curve is the curve defined by Eq.(1).*

Let us now consider a periodic corrugated surface whose cross section (in the xy-plane) is depicted in Fig. 3. It has the same depth $h = 4$ mm as the grounded slab, the same permittivity, but metallic ridges have been added. Figure 4 gives the phases of the reflection coefficients. The dashed curve still represents Eq. (1) and facilitates a comparison with the grounded slab. We can see that the $R_{TE}$ coefficient stays very similar to that of the grounded slab, contrary to the $R_{TM}$ coefficient. Note that using our numerical codes specially developed for these lamellar gratings [13-15], such graphs are computed in some minutes.

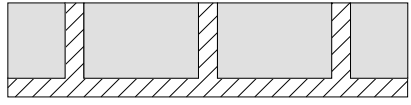

*Fig. 3: Cross section of a corrugated surface. The metallic (infinitely conducting) part is dashed. The grayed area is filled with dielectric with $\varepsilon = 2.2$. The period is 10 mm, and the width of the ridges is 2.6 mm.*

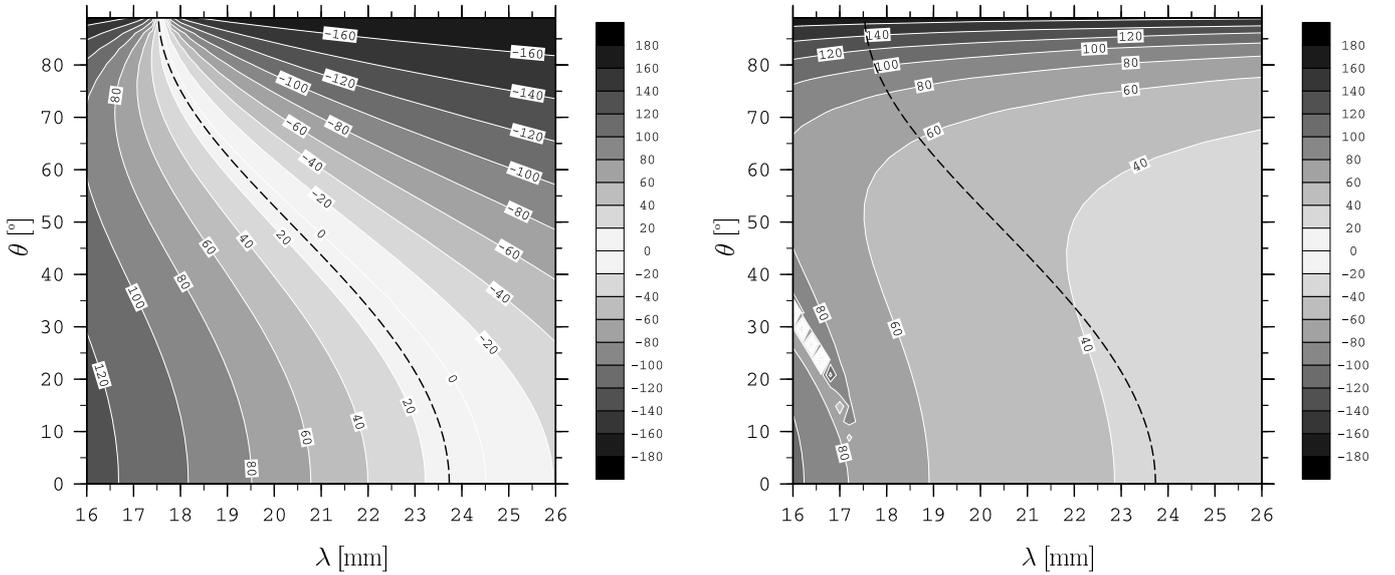

*Fig. 4: Phases of $R_{TE}$ (left) and $R_{TM}$ (right) for the corrugated surface of Fig. 3. The dashed curve is the curve defined by Eq.(1).*

We now consider a dielectric slab covered with strips (Fig. 5). The parameters have been chosen in order to get $R_{TE} = +1$ for the same values $\theta \approx 90°$ and $\lambda \approx 17.5$ mm as in the previous cases. Figure 6 gives the phases of the reflection coefficients of this surface. Clearly, this surface behaves like a hard surface ($R_{TE} \approx +1$ and $R_{TM} \approx +1$) along the curve $\arg(R_{TE}) = 0$. The bandwidth is very narrow near grazing incidences, and becomes larger near normal incidences.

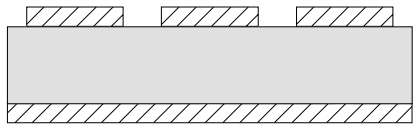

*Fig. 5: Cross section of a dielectric slab covered with strips (schematic, not scaled). The metallic (infinitely conducting) part is dashed. The grayed slab has a thickness $h = 0.8$ mm and a permittivity $\varepsilon = 2.2$. The period is 10 mm, the width of the strips is 7.4 mm, and their thickness is 0.03 mm (not critical).*

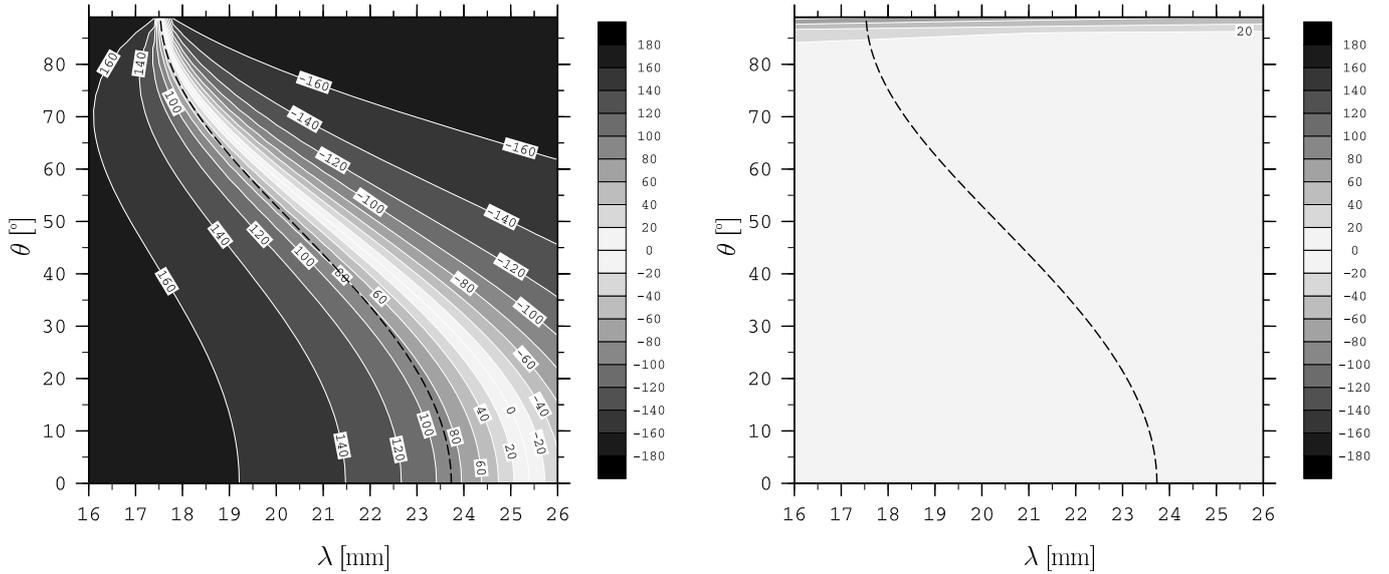

*Fig. 6: Phases of $R_{TE}$ (left) and $R_{TM}$ (right) for the surface of Fig. 5. The dashed curve is still the same as in Figs. 2 and 4.*

Figure 7 shows field maps of the electric field in the conditions that are suitable for the propagation of the TEM mode. In order to compare the last two structures, the same scale is used in both graphs, and the top of the two surfaces are placed at $y = 0$. As expected, the field amplitude tends towards a value equal to 2 (in phase addition of the incident and the reflected propagating plane waves) at some distance from the top of the structure. But these graphs also show the behavior of the near field and the role of the evanescent waves. They explain the apparent contradictory situation mentioned above: the evanescent waves make the transition between the boundary conditions enforced by the metallic parts and the approximate problem where only the propagative waves are considered. We can see that, even if the dielectric slab covered with strips is a thinner structure than the corrugated one, this advantage is attenuated by the fact that the field is more resonant. If we define an "electromagnetic height" of the structure by the distance between the bottom of the structure and a plane where the electric field is more or less uniform, this electromagnetic height is roughly equal to 6 mm for the corrugated surface and 5 mm for the slab covered with strips.

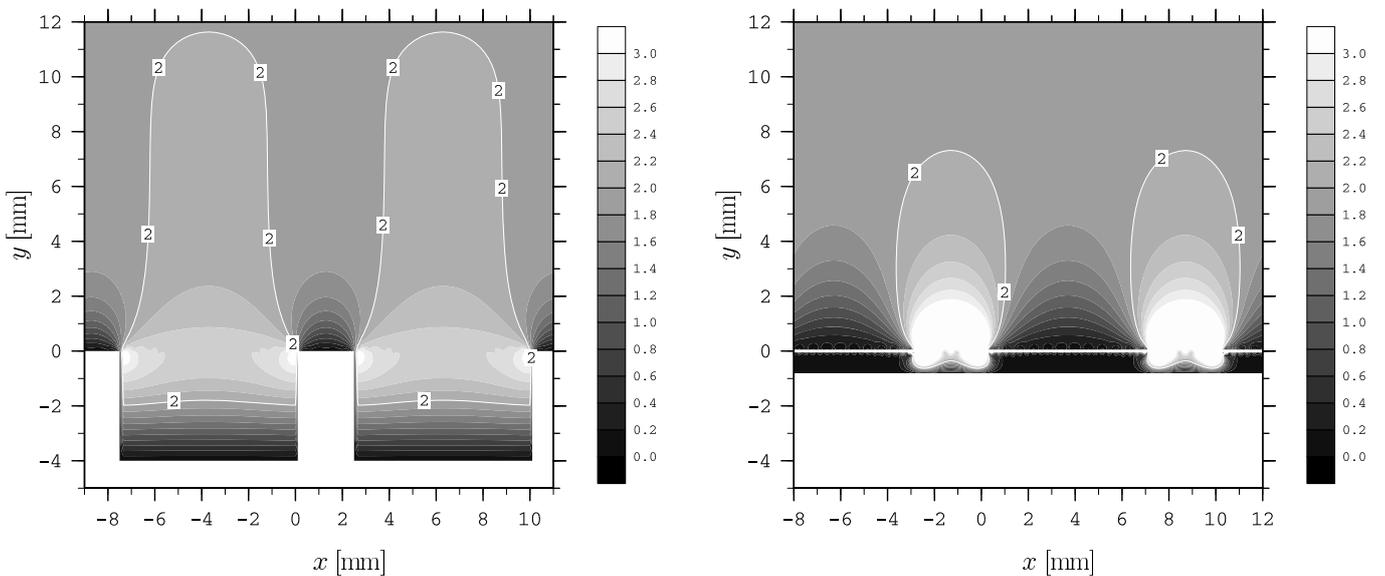

*Fig. 7: Modulus of the electric field in TE polarization and for grazing incidence $\theta = 89°$. Left: corrugated surface at $\lambda = 17.6$ mm. Right: dielectric slab covered with strips at $\lambda = 17.58$ mm. The metallic parts are represented in white.*

### 4. From the reflection coefficients of the surfaces to the dispersion curves of quasi-TEM guided modes

It has been seen in the previous section how PMC or hard surfaces permit the propagation of a TEM mode in a rectangular waveguide. But this approach also shows that this TEM propagation can only exist for a very particular wavelength. In the present section, we investigate how a quasi-TEM mode can propagate in rectangular waveguides for various wavelengths.

Let us ignore the evanescent waves which vanish rapidly when going away from the surface, and consider only propagating TE waves with an incidence θ (Fig. 8). We obtain a propagating mode inside the structure when the following phase matching condition is realized, which expresses that the wave is in phase with itself after 2 reflections on the surfaces:

$$2a\cos\theta + \frac{\lambda}{\pi}\arg(R_{TE}) = m\,\lambda \qquad (2)$$

where $a$ is the distance between the 2 surfaces, and $m$ an integer. Since the electric field is perpendicular to the Fig. 8, the solution will stay valid for a closed waveguide if we use appropriate surfaces for the two others walls (PEC surfaces for instance, but hard surfaces are another possibility).

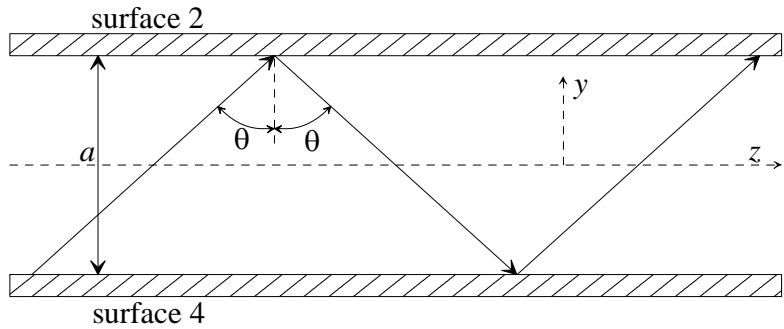

*Fig. 8: Propagation of a guided mode. The electric field is perpendicular to the figure.*

From the computations done in the previous section, we know the phase of $R_{TE}$ as a function of λ and θ. Let us denote by $\arg(R_{TE}) = f(\lambda,\theta)$ this relationship. We can now solve the implicit system

$$\begin{cases} \arg(R_{TE}) = f(\lambda,\theta) \\ 2a\cos\theta + \dfrac{\lambda}{\pi}\arg(R_{TE}) = m\,\lambda \end{cases} \qquad (3)$$

in order to get a relationship between λ and θ for each value of $m$. Figure 9 gives the solution of this implicit system (the dispersion curves of the guided modes) in the case where surfaces 2 and 4 are similar to that of Fig. 5, i.e. a dielectric slab covered with strips. The parameters taken here for these surfaces are: thickness $h = 0.25$ cm, permittivity $\varepsilon = 5$, period equal to 0.3 cm, width of the strips equal to 0.2 cm, and thickness of the strips equal to 0.03 mm. On the dispersion curves, we have pointed out the condition where the TEM mode exists (θ = 90° and wavelength $\lambda_1$). For another wavelength such as $\lambda_2$, we can verify that the phase matching condition (2) is verified, from the value of $\arg(R_{TE})$ read on Fig. 9 (left).

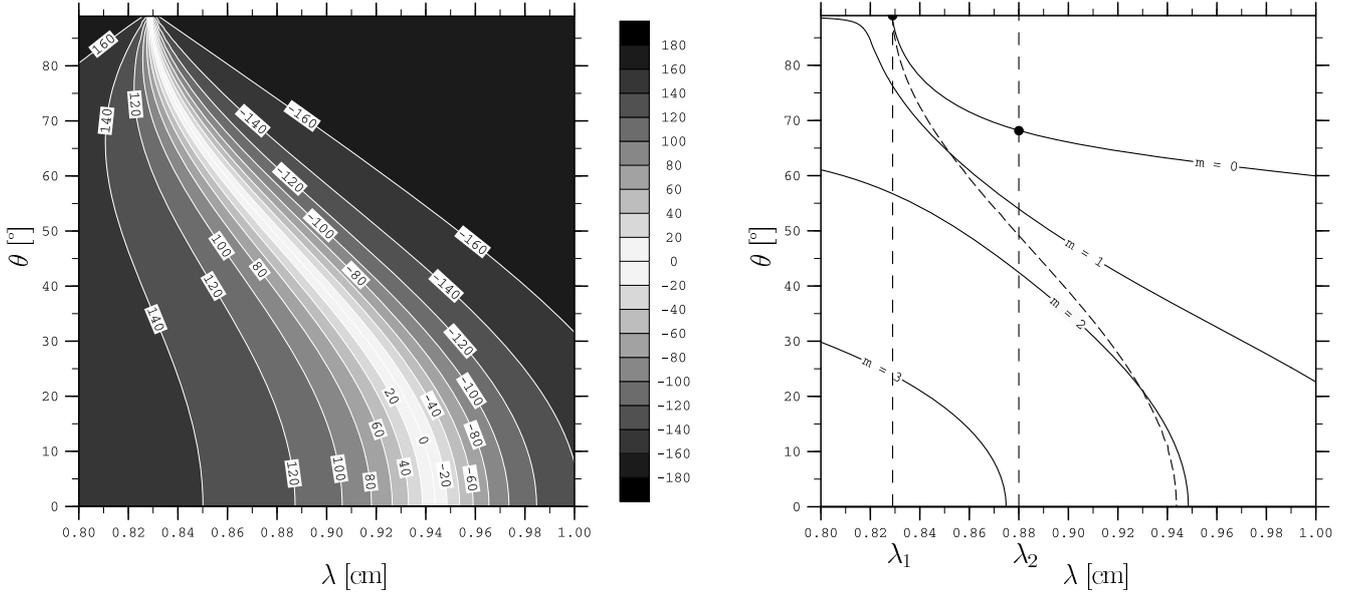

*Fig. 9. Left: phase of $R_{TE}$. Right: dispersion curves of the guided modes $m = 0,1,2,3$, for a guide with dimension $a = 1$ cm. The superposed dashed curve is the curve $\arg(R_{TE}) = 0$. The wavelengths are in cm.*

When departing from the grazing incidence $\theta \approx 90°$, the uniformity of the field map deteriorates. From Fig. 8, we see that the field in the waveguide is the superposition of two plane waves propagating in directions which make an angle $(\pi/2) - \theta$ with the z-axis. It can easily be deduced that in a section of the waveguide, the interference between these plane waves will give a pattern with fringes parallel to the *x*-axis and with an interfringe (distance between fringes along the *y*-axis) equal to

$$i = \frac{\lambda}{2\cos\theta} \ . \tag{4}$$

For example, if we consider the propagation at $\lambda_2 = 0.88$ cm and $\theta = 70°$ in Fig. 9, the interfringe is 1.29 cm, and since the waveguide dimension is $a = 1$ cm, we can deduce that the uniformity of the mode in a cross-section of the waveguide will be notably degraded.

In order to check these results, let us consider the waveguide shown in Fig. 10. The modes of this waveguide have been studied with the help of a numerical code recently developed in our laboratory and based on the fictitious sources method [16,17]. The mode finding reduces to an eigenvalues problem, and to the search of null eigenvalues. It is not our purpose to detail the method here, and we will just say that the dispersion curves are obtained by the study of the modulus of a determinant, which gives the gray scale map represented on Fig. 11. The upper area of this map is the domain of grazing incidences, where this method is not sensitive and gives no information about the guided modes. But for smaller incidences, we can see that it retrieves the expected modes, particularly for $m = 0$, 1, and 2. We also see the existence of some other modes that result from the finite size of the waveguide in the *x* direction. Those modes have fields that vary along the *x*-axis, and are not taken into account in our analysis focused on modes with a quasi-uniform field map inside the cross section (our analysis based on plane waves implies that the field is invariant along the *x*-axis).

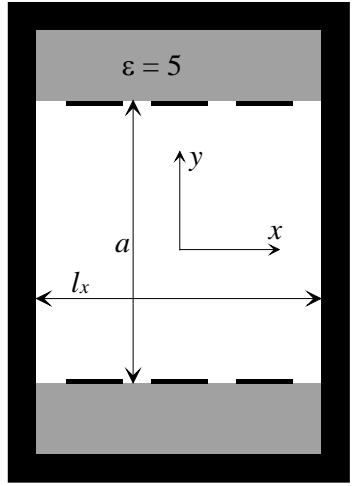

*Fig. 10. Cross-section of the waveguide. Black represents PEC metallic parts, gray represents dielectric.*
$a = l_x = 1 \,\text{cm}$.

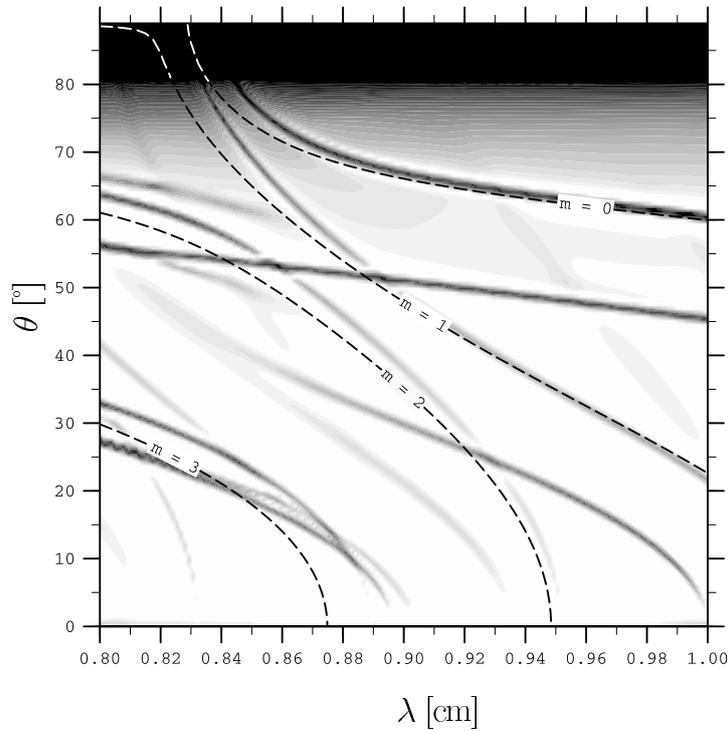

*Fig. 11. Dispersion curves of the modes of the waveguide shown in Fig. 10, computed by the fictitious sources method (gray scale map). The superposed curves are those deduced from Eq. (3).*

## 5. Study of the quasi-TEM modes inside a waveguide

Let us now turn to a practical application, using a square waveguide with four hard walls that allow the propagation of a dual-polarized quasi-TEM mode. This situation can be interesting for devices able to work with two orthogonal polarizations such as resonant cavities antennas for example. These surfaces are similar to that of Fig. 5, i.e. a dielectric slab covered with strips. Their parameters are now: dielectric thickness $h = 0.25$ cm, permittivity $\varepsilon = 5$, period equal to 0.5 cm, width of the strips equal to 0.3 cm, and thickness of the strips equal to 0.03 mm.

For clarity, it is convenient to draw the dispersion curves of the modes in a more conventional way. We can transform the coordinate system $(\lambda, \theta)$ used up to now into the coordinates $(f, k_z)$ using simple transformations. The frequency $f$ is simply $c/\lambda$. The propagation constant $k_z$ of the mode along the $z$ axis is deduced from Fig. 8: $k_z = (2\pi/\lambda)\sin\theta$. Figure 12 shows $\arg(R_{TE})$ corresponding to the present surface in these two coordinate systems. Note that on the conventional graph in variables $(f, k_z)$, the TEM modes, which correspond to values of $\theta \approx 90°$, are located near the line $k_z = 2\pi / \lambda = 2\pi f / c$. In this case, the TEM mode corresponds to $\theta = 90°$, $\lambda = 26.6$ mm, $f = 11.3$ GHz, $k_z = 236$ m$^{-1}$.

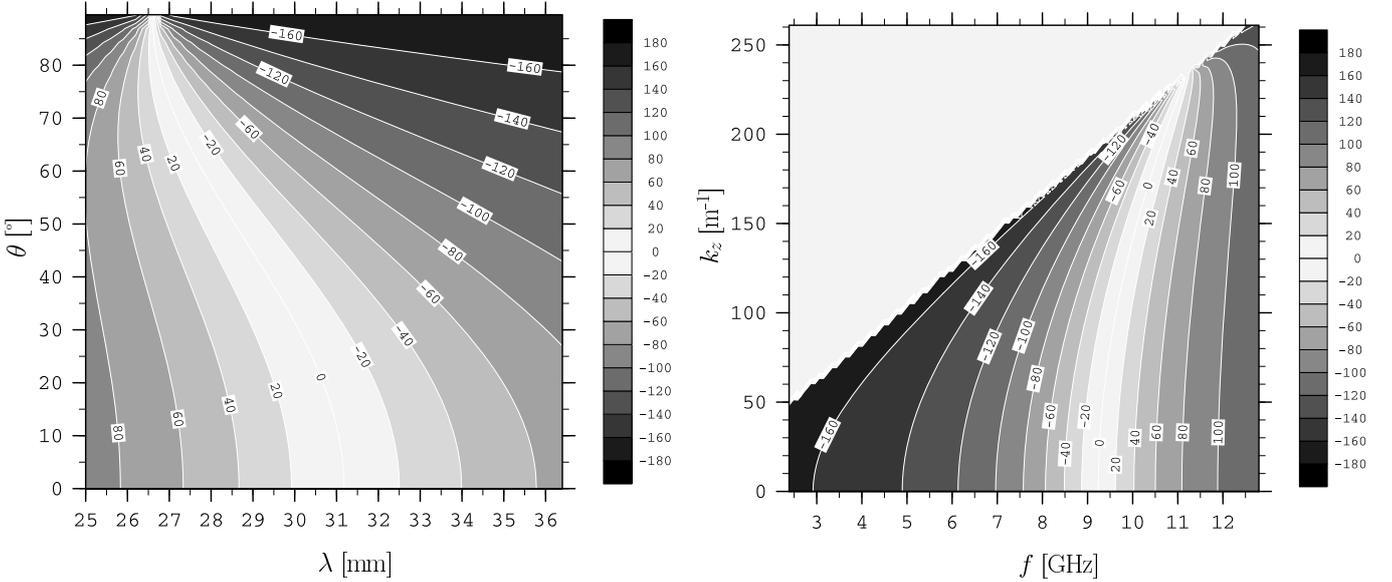

*Fig. 12. Phase of $R_{TE}$ for the hard surface used in section 5, represented in the two coordinate systems $(\lambda, \theta)$ and $(f, k_z)$*

The waveguide cross section is represented on Fig. 13. Its dimensions are: $a = l_x = 47$ mm. From the knowledge of $R_{TE}$ and the value of $a$, we deduce the dispersion curves of the guided modes that are x-invariant. These dispersion curves are given on Fig. 14, in the two coordinate systems.

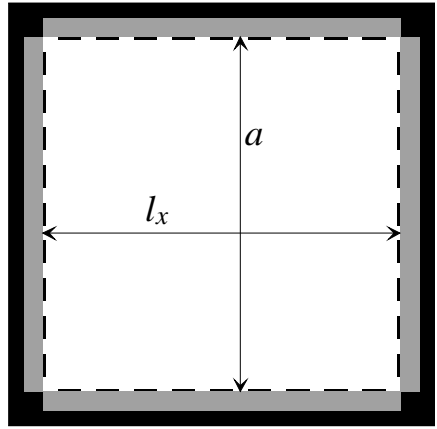

*Fig. 13. Waveguide cross section.*

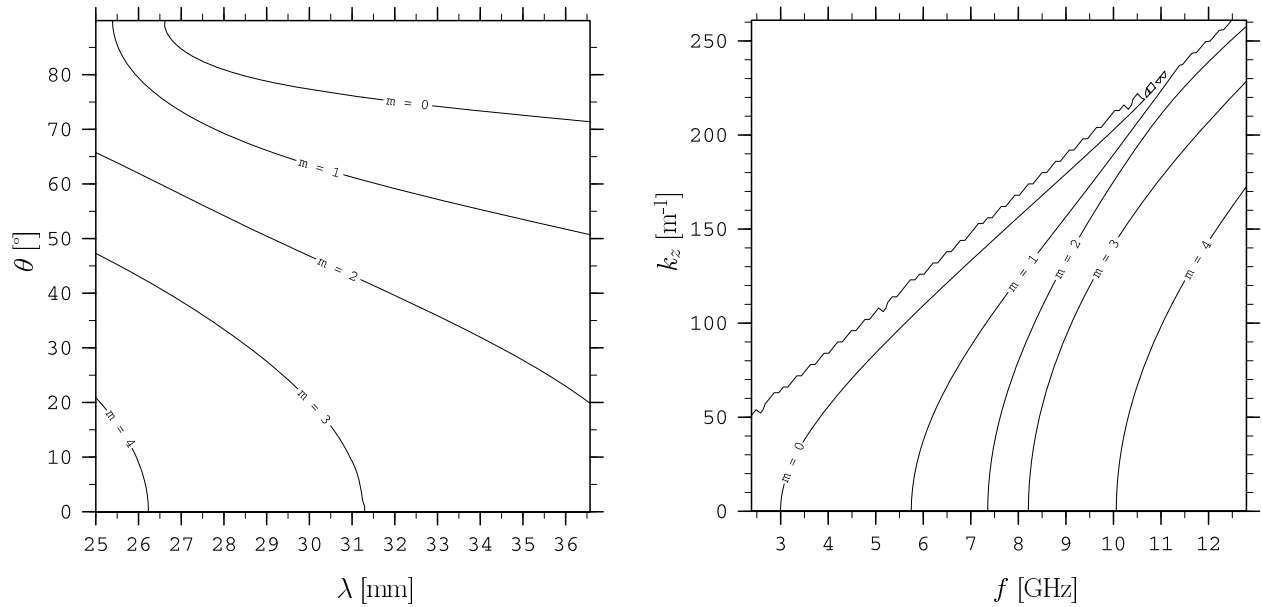

*Fig. 14. Dispersion curves of the x-invariant guided modes in the waveguide of Fig. 13. The imprecision of the line $k_z = 2\pi f / c$ is due to the implicit solver used to get these curves.*

In order to check the uniformity of the field, we have computed the mode corresponding to $m = 0$ for different frequencies. The computations have been made with the help of a commercial solver, and only a quarter of the cross-section is represented on the maps of Fig. 15.

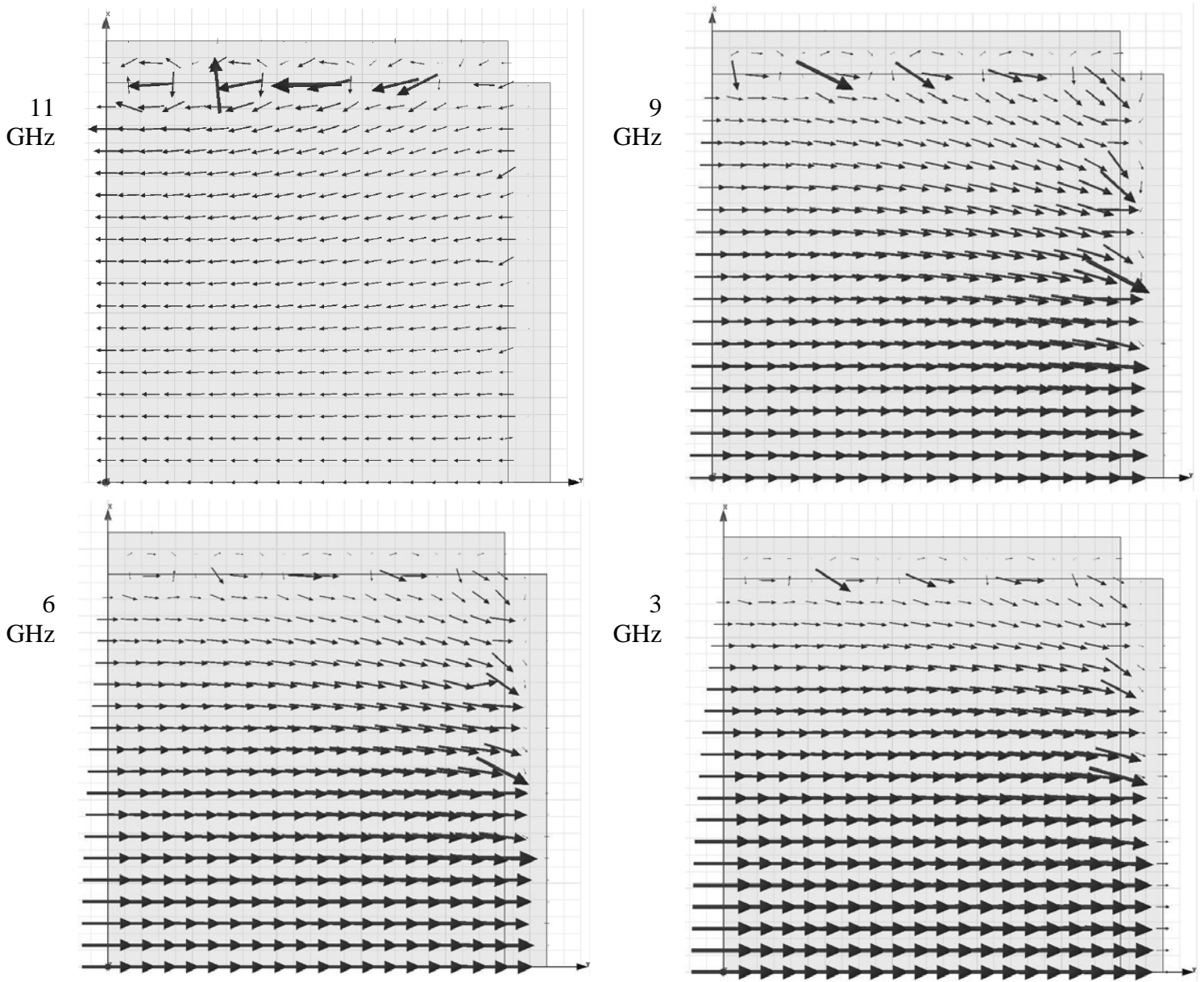

*Fig. 15. Electric field distribution in a quarter of the cross-section for various frequencies.*

We can check the interfringe formula given by Eq. (4). At 9 GHz, $\lambda = 33.3$ mm, Fig. 14 (left) gives $\theta = 75°$, and finally Eq. (4) gives $i = 64$ mm. At the mode cutoff, $f = 3$ GHz, $\lambda = 100$ mm, Fig. 14 (right) gives $\theta \approx 0$, and $i = \lambda/2 = 50$ mm. In this case, the field map presents a maximum in the centre of the waveguide, and a marked minimum on the upper and lower walls. In any case, the field uniformity along the *x*-axis is verified. The evanescent waves due to the strips allow the field to behave nicely in the vicinity of each of the 4 walls. Finally, it can be observed that the mode keeps a good uniformity in an interesting range of frequencies.

**Conclusion**

We have presented a study of quasi-TEM modes inside a rectangular waveguide based on the use of reflection coefficients of the surfaces that make the walls of the waveguide. By this way, all the computations that permit to get the dispersion curves of the quasi-TEM modes can be made quickly. Of course, the detailed study of the modes inside the waveguide needs more sophisticated and time consuming codes. This study allowed us to get a clear insight upon these modes and can be useful for the design of such structures. In particular, there exist many surfaces able to guide quasi-TEM modes, and their design is not obvious, due to the multiple geometric and electromagnetic parameters that are involved. With these results in mind, it could be simpler to optimize the structures in order to obtain given specifications for the guided modes: optimization of the bandwidth, choice between corrugated or strip covered surfaces (the latter present stronger local fields and this can be a drawback for some structures when taking losses into account), and so on.